\documentclass[onecolumn,amsmath,amssymb,nofootinbib,letterpaper,11pt]{article}
\usepackage{jheppub}
\usepackage{ifpdf}
\usepackage[utf8]{inputenc}
\usepackage[T1]{fontenc}

\input{epsf}
\usepackage{epsfig}
\usepackage{epstopdf}
\usepackage{arcs}

\usepackage{graphicx}
\usepackage{subfigure}

\usepackage{tensor}
\usepackage{braket}
\usepackage{float}
\usepackage[dvipsnames]{xcolor}
\usepackage{amsmath}
\usepackage{morefloats}
\newcommand{\be}{\begin{equation}}
\newcommand{\ee}{\end{equation}}

\usepackage{tikz}
\usetikzlibrary{calc}   
\usetikzlibrary{topaths}
\usepackage{hyperref}
\hypersetup{
	colorlinks =WildStrawberry,
	linkcolor =red,
	pdftitle={Mimetic Inflation},
	citecolor=NavyBlue,
	filecolor=WildStrawberry,
	urlcolor=WildStrawberry,
}

\usepackage{color}

\usepackage{array}
\usepackage{bm}

\usepackage[toc]{appendix}
\usepackage{color,soul}
\usepackage{datetime}
\newcommand{\bea}{\begin{eqnarray}}
\newcommand{\eea}{\end{eqnarray}}
\newcommand{\ba}{\begin{eqnarray}}
\newcommand{\ea}{\end{eqnarray}}

\newcommand{\beq}{\begin{equation}}
\newcommand{\eeq}{\end{equation}}
\newcommand{\beqa}{\begin{eqnarray}}
\newcommand{\eeqa}{\end{eqnarray}}
\newcommand{\beqar}{\begin{eqnarray*}}
\newcommand{\eeqar}{\end{eqnarray*}}

\renewcommand{\c}{$c$}

\renewcommand{\href}[2]{#2}

\title{Universal criticality of thermodynamic geometry for boundary conformal field theories in gauge/gravity duality}

\author{Morteza Rafiee $^{a}$\footnote{m.rafiee@shahroodut.ac.ir}, }
\author{Seyed Ali Hosseini Mansoori $^{a}$\footnote{shosseini@shahroodut.ac.ir},}
\author{Shao-Wen Wei $^{b,c}$\footnote{weishw@lzu.edu.cn},}
\author{Robert B. Mann $^{d,e}$\footnote{rbmann@uwaterloo.ca},}

\affiliation[a]{Faculty of Physics, Shahrood University of Technology, P.O. Box 3619995161 Shahrood, Iran}
\affiliation[b]{Institute of Theoretical and Physics Research Center of Gravitation,
	Lanzhou University,\\ Lanzhou 730000, People’s Republic of China}
	\affiliation[c]{Institute of Theoretical Physics and Research Center of Gravitation, Lanzhou University, Lanzhou 730000, China}
	\affiliation[d]{Perimeter Institute, 31 Caroline St. N. Waterloo, Ontario, N2L 2Y5, Canada}
	\affiliation[e]{Department of Physics and Astronomy, University of Waterloo, Waterloo, Ontario, N2L 3G1, Canada}

\vspace{0.1cm}

\abstract{According to more recent AdS/CFT interpretation \cite{Karch:2015rpa}, in which varying
	cosmological constant $\Lambda$ in the bulk corresponds to varying the curvature radius
	governing the space on which the field theory resides, we study the criticality of thermodynamic curvatures for thermal boundary conformal field theories (CFT) that are dual to $d$-dimensional charged anti-de Sitter (AdS) black holes, embedded in $D$-dimensional M-theory/superstring inspired models having $AdS_{d}\times \mathbb{S}^{d+k}$
	spacetime with $D=2d+k$. Analogous with  criticality features acquired for charged AdS black holes in the bulk \cite{HosseiniMansoori:2020jrx}, the normalized intrinsic curvature $R_N$ and extrinsic curvature $K_N$ of the boundary CFT has critical exponents 2 and 1, respectively. In this respect, the universal amplitude of $R_Nt^2$ is $\frac{1}{2}$ and $K_Nt$ is $-\frac{1}{2}$ when $t\rightarrow0^-$, whereas $R_Nt^2\approx \frac{1}{8}$ and $K_Nt\approx\frac{1}{4}$ in the limit $t\rightarrow0^+$ in which  $t=T/T_c-1$ is the temperature parameter with the critical temperature, $T_{c}$. Interestingly,  these critical amplitudes are independent of the number of thermal CFT dimensions and are remarkably similar to one given for higher dimensional charged AdS black holes in the bulk. }

\preprint{
}

\begin{document}
	\maketitle

	\section{Introduction}
	
Quite some time ago, the existence of   small-large black hole phase transitions of charged and rotating AdS black holes, along with their similarity with the liquid-gas phase transition of  Van der Waals (VdW) type were pointed out
\cite{Chamblin:1999tk, Chamblin:1999hg}. A complete identification between the thermodynamic parameters of a charged black hole and a Van der Waals system was   made some years later in the framework of  extended black hole thermodynamics, in which the cosmological constant was treated as thermodynamic
	pressure and its conjugate quantity as a thermodynamic volume of the black hole \cite{Kubiznak:2012wp}. 
This in turn led to a broad range of 	theoretical developments in the study of thermodynamics and phase transitions of AdS black holes in the extended phase space \cite{Altamirano:2013ane,Altamirano:2013uqa,Altamirano:2014tva,Wei:2012ui,Cai:2013qga, Sherkatghanad:2014hda,Dolan:2014vba}, in a subdiscipline generally referred to as black hole chemistry 	\cite{kubizvnak2017black}.
	
	In the framework of gauge/gravity duality \cite{Maldacena:1997re}, it is therefore of interest to ask
	whether the interpretation of the dynamical cosmological constant as pressure is
	applicable to the CFT side?
Early attempts  to address this question
	\cite{Johnson:2014yja,Kastor:2014dra,Dolan:2014cja} proposed that varying pressure, or cosmological constant $\Lambda$, 
	in the bulk corresponds to varying the number of colors, $N$, in the boundary
	Yang–Mills theory, with thermodynamic volume  interpreted in the boundary
	field theory as an associated chemical potential $\mu$ for color.  For  example,
	chemical potentials corresponding to neutral and
	charged black holes in $AdS_{5}\times \mathbb S^{5}$ have been  explicitly obtained by
       $N^2$ (instead of $N$) as the number of degrees of freedom
	of the $\mathcal{N}=4$ SU$(N)$ Yang–Mills theory in the large $N$ limit
       \cite{Zhang:2014uoa,Zhang:2015ova,Maity:2015ida,Mahish:2020gwg,Chabab:2015ytz,Wei:2017icx}. In this approach
  it is worth emphasizing that varying $\Lambda$ in the AdS bulk is
	equivalent to changing the boundary field theory.
	
	An alternative approach followed here is based on a different interpretation,
	which proposes that $N$ should be kept fixed, so that field theory remains the
	same. In this approach, varying $\Lambda$ in the bulk is equivalent to varying
	the curvature radius governing the space on which the field theory resides
	\cite{Karch:2015rpa}. In other words, in order to keep $N$ fixed, we  retain the
	standard holographic relation 	\begin{equation}\label{Ieq1}
	N^{p}\sim \frac{L^{d-2}}{G_{d}}
	\end{equation}
	between $N$,  the AdS radius $L$, and the $d$-dimensional
	gravitational constant $G_{d}$. To keep the number of colors $N$ fixed we must vary $G_{d}$ and  $L$ whilst keeping the combination
	$L^{d-2}/G_{d}$ fixed. Note that the  power $p$ depends on a theory one considers. For
	example, for a gauge theory like $\mathcal{N}=4$ SuperYang-Mills,  $p = 2$,
	whereas for M-theory in $d=11$ compactified on $AdS_{4} \times \mathbb S^{7}$
	with a three-dimensional CFT,   $p=3/2$ \cite{Maldacena:1997re}.
	In other words, this interpretation proposes that varying the cosmological constant
	(or equivalently the AdS length scale) in the bulk corresponds to varying the
	spatial volume  in the dual field theory, which is completely determined by the radius of
	the sphere on which the field theory resides \cite{Karch:2015rpa}.  From a 
	thermodynamic point of view, the conjugate variable associated with this volume
	is pressure, and one is thus led to study the extended thermodynamics of the CFT
	\cite{Dolan:2016jjc}.

Using this latter interpretation of the boundary extended thermodynamics, in this paper we carry out
 the first investigation of the thermodynamics of thermal boundary field theories  by making use of 
 a new formalism for the geometry of the extended thermodynamic phase space \cite{Mansoori:2013pna,HosseiniMansoori:2019jcs,Mansoori:2016jer}  that we shall describe below.  We shall concentrate
 on field theories that are  dual to charged AdS black holes in the bulk. Such black holes can be
embedded in $D$-dimensional M-theory/superstring inspired models having $AdS_{d} \times \mathbb S^{d+k}$
space-time with $D = 2d + k$.  A key aim of our paper is to investigate the universality of the thermodynamic geometry for boundary field theories ($CFT_{d-1}$) that  are dual to $d$-dimensional charged AdS black holes ($AdS_{d}$) in the bulk. We now go on to outline our approach.

It has become clear in recent  years that  the geometry of thermodynamic  phase space, as described using the Ruppeiner metric,  provides us with a useful tool for studying black hole phase transitions and criticality \cite{reff8,reff9,reffff9,reff12,reff10,reff4,reff5}.  In the first holographic interpretation of black hole chemistry, in which the number of colors is varied \cite{Johnson:2014yja,Kastor:2014dra,Dolan:2014cja}, the Ruppeiner geometry
geometry has been extensively applied toward the study of phase transitions and to probe the nature of microscopic interactions in black holes \cite{Zhang:2014uoa,Zhang:2015ova,Mahish:2020gwg, Chabab:2015ytz,Wei:2017icx}.

Recently a new formalism of thermodynamic geometry (called NTG)  \cite{Mansoori:2013pna,HosseiniMansoori:2019jcs,Mansoori:2016jer} was developed that yields a one-to-one correspondence between phase transition points and singularities of the corresponding scalar curvature.  This useful feature overcomes a key limitation \cite{reff10,Sarkar:2006tg} of the standard Ruppeiner approach  \cite{HosseiniMansoori:2020jrx} and has yielded other interesting results.
Specifically, the intrinsic (scalar) and extrinsic curvatures of the thermodynamic geometry were found to exhibit universal behavior near critical point for $d$-dimensional charged AdS black holes in the bulk. These black holes, which have behaviour qualitatively similar to that of a Van der Waals fluid in the context of the black hole chemistry \cite{Kubiznak:2012wp}, 
always possess  critical exponents 2 and 1 for their respective (normalized) intrinsic and extrinsic NTG curvatures near the critical point  in any space-time dimension \cite{HosseiniMansoori:2020jrx}. In addition, their critical amplitudes were found to be	\begin{eqnarray}\label{Adscriticality}
	R_{N} t^2  \approx \bigg\{ \begin{matrix}
	-\frac{1}{2} & \text{for} & t>0,\\
	-\frac{1}{8} & \text{for} & t<0,
	\end{matrix} \hspace{0.5cm} \text{and} \hspace{0.5cm} K_{N} t  \approx \bigg\{ \begin{matrix}
	-\frac{1}{2} & \text{for} & t>0,\\
	-\frac{1}{4 } & \text{for} & t<0.
	\end{matrix}
	\end{eqnarray}
where the temperature parameter $t=T/T_{c}-1$,  with $T_{c}$  the critical temperature. Interestingly, such  amplitudes are not dependent on the number of spacetime dimensions, indicative of their universality  \cite{HosseiniMansoori:2020jrx}. 
	
As stated above, we are interested as to whether or not such universality  holds for boundary field theories ($CFT_{d-1}$) that  are dual to bulk charged AdS black holes in $d$-dimensions, working in the interpretation where $N$ is kept constant whilst the  cosmological constant is variable \cite{Karch:2015rpa,Cong:2021fnf,Visser:2021eqk}.  We shall see that there is indeed universality in the dual boundary theory, though the numerical details differ slightly.
 	
	Our paper is organized as follows. In Section \ref{CFTSEC}, we investigate the behaviour and criticality of some thermodynamic quantities such as  heat capacities and compressibilities  of the $(d-1)$ dimensional boundary CFT that is dual to a $d$- dimensional AdS black hole embedded in
	$D$-dimensional superstring/M-theory inspired models. In Section \ref{TCG}, the NTG geometry is employed in investigating phase transitions and critical behavior of boundary field theories. Remarkably, the intrinsic and extrinsic NTG curvatures  reveal universal behaviour around the critical point for boundary CFTs. Our conclusions are presented in Section \ref{con}.    
	
	\section{The critical behavior of the boundary conformal field theory}\label{CFTSEC}
	
We are interested in studying the behaviour and criticality of some thermodynamic quantities of a $(d-1)$ dimensional boundary CFT that is dual to a $d$- dimensional AdS black hole embedded in $D$-dimensional superstring/M-theory inspired models.  	
	To be more precise, a $d$-dimensional AdS black hole can be embedded in
	D-dimensional superstring/M-theory inspired models having $AdS_{d} \times \mathbb{S}^{d+k}$ space-time, where
	$D=2d+k$. These black hole solutions can be related to $N$ coincident $(d-2)$-branes that are  assumed to move in such higher dimensional models and are labeled by a triplet $(D; d; k)$, where
	$k$ is associated with the internal space, i.e., the $\mathbb{S}^{d+k}$ sphere \cite{Chamblin:1999tk}. For example, the triplet $(D,d,k)=(11,7,-3)$ is associated with  the compactification of M-theory on the 
	sphere $\mathbb{S}^4$  with   $M5$-branes (see   \cite{Belhaj:2021aqq} for more details). One can also consider 
	eleven-dimensional M-theory, with $ AdS_{4}\times \mathbb{S}^{7}$ space-time in the presence of $N$ coincident $M2$-branes,  which is characterised by	the triplet $(D,d,k)=(11,4,3)$ \cite{Dabholkar:2014wpa}. In addition,  the ten-dimensional type IIB superstring theory with the $AdS_{5}\times \mathbb{S}^{5}$ space-time in the presence of $D3$-branes is realized by the triplet $(D,d,k)=(10,5,0)$ \cite{Witten:1998qj,Chamblin:1999tk}.
	
	Furthermore, the gravitational constant $G_d$ in such a $d$-dimensional AdS black hole is associated to  one corresponding to $D$-dimensional superstring/M-theory inspired models \cite{Gubser:1998nz}. Since the 
	$d$-dimensional AdS black hole is obtained from compactification of 
	the $D$-dimensional theory on the $\mathbb{S}^{d+k}$ sphere of radius $L$, the $d$-dimensional Newton constant $G_{d}$ is descended from its $D$-dimensional counterpart $G_{D=2d+k}$ via
	\begin{equation}
	G_{d}=\frac{G_{2d + k}}{\text{Vol}\left( \mathbb{S}^{d+k} \right)}=\frac{G_{2d + k}}{ \Omega_{d+k} \, L^{d+k}},
	\label{Geq}
	\end{equation}
	where $\Omega_{d+k}=2 \pi^{(d+k+1)/2}/\Gamma(\frac{d+k+1}{2}) $ is the volume of a unit $(d+k)$ sphere.
	The AdS radius $L$ is also related to the brane number $N$ via
	\begin{equation}
	L^{2(d-1)+k}=2^{-\left( \frac{d\left(4-d\right)+3}{2}  \right)} \, \pi^{7\left( k+2(d-5) \right)-4} \, N^{\frac{d-1}{2}} \ell_{p}^{ \, {2(d-1)+k}},
	\label{Leq}
	\end{equation}
	where the  $D$-dimensional Planck length $l_{p}$ is associated with
	Newton’s constant $G_{2d+k}$ via $l_{p}^{2(d-1)+k}=\hbar G_{2d+k}$. Clearly,  varying the AdS radius $L$ at fixed $G$ leads to
	the variation of the color number $N$ on the field theory side 	\cite{Johnson:2014yja,Kastor:2014dra,Dolan:2014cja}.
	On the other hand, combining the two
	preceding relations, we arrive at
	\begin{equation}\label{gd}
	\frac{1}{4 \pi \hbar G_d}=\gamma\frac{N^{\frac{d-1}{2}}}{\mathcal{V}}
	\end{equation}
	which is consistent with Eq. \eqref{Ieq1}, where $\gamma=2^{\frac{1}{2} ((d-4) d-7)} \pi^{14 d+7 k-75} \Omega_{d+k} \Omega_{d-2} $ is a constant number and $\mathcal{V}=\Omega_{d-2} L^{d-2}$ can be
	interpreted as the spatial volume of the boundary conformal field theory by $(d-1)$ dimensions.
	It
	follows that varying the volume of the CFT, keeping $N^{d-1/2}$ fixed, is
	completely equivalent to varying $G_d$, or the $d$-dimensional Planck length \cite{Karch:2015rpa}. From this latter perspective  we study the 
	thermodynamics of the boundary conformal field theory. Indeed, we assume that $N^{d-1/2}$ should be kept fixed, so that varying $\Lambda$ in the
	bulk has the more natural consequence of changing the volume of the boundary conformal field theory
	\cite{Karch:2015rpa}.  
Using this interpretation, a generalized Smarr
	relation was obtained by taking extended black hole thermodynamics
	of the boundary field theory \cite{Sinamuli:2017rhp}. Furthermore a precise boundary
	description of extended black hole thermodynamics,  in which  varying $\Lambda$ only
	corresponds to varying central charge ($C$) (instead of  varying the volume of the boundary field theory)
	was recently proposed  \cite{Cong:2021fnf,Visser:2021eqk}. In addition, by including the variation of the gravitational constant $G$ in the first law, a new definition for thermodynamic black	hole volume can be defined in the bulk, whilst the
	CFT remains unchanged (fixed C) on the boundary \cite{Cong:2021fnf}.

 The static and spherically symmetric $d$-dimensional charged AdS black hole solution in  Einstein-Maxwell-AdS gravity is given by \cite{Chamblin:1999tk}
	\begin{eqnarray}\label{metric1}
		ds^2&=&-f(r)dt^2+\frac{dr^2}{f(r)}+r^2d^2\Omega_{d-2},	\\
	\nonumber 	f(r)&=&1-\frac{m}{r^{D-3}}+\frac{q^2}{r^{2(D-3)}}+\frac{r^2}{L^2},	
	\end{eqnarray}
where	
\begin{equation}\label{mq1}
		M=\frac{(D-2) \Omega_{d-2} m}{16 \pi G_D} \qquad   Q =\sqrt{\frac{(d-2)(d-3)}{2}} \Omega_{d-2} q
	\end{equation}
	respectively relate the parameters $m$ and $q$  to the black hole mass $M$ and
	charge $Q$ \cite{Chamblin:1999tk}.
Imposing the condition $f(r_h)=0$ at the event horizon $r_{h}$ yields
	\begin{eqnarray}\label{mass-entropy}
		M=\frac{(d-2)\Omega_{d-2}}{16 \pi G_d
		}\left(\frac{r_h^{d-1}}{L^2}+r_h^{d-3}+\frac{q^2}{r_h^{d-3}}\right), 
	\end{eqnarray} 
expressing  $M$ in terms of $r_{h}$ and $q$. 
	Moreover, the Bekenstein-Hawking entropy is 
	\begin{equation}
	S=\frac{A_{h}}{4 \hbar G_{d}}=\frac{\Omega_{d-2}}{4}\frac{r_h^{d-2}}{\hbar G_d}.
	\end{equation}
	In the context of the AdS/CFT dictionary, black hole thermodynamics can be understood in terms of the fundamental degrees of freedom of a thermal
	quantum field theory and vice versa. Therefore, one can map the black hole mass $M$ to the
	internal energy $U$ of the large $N$ Yang-Mills theory at the boundary, and the temperature $T$ and entropy
	$S$ of the black hole to those of the boundary field theory
	\cite{Chamblin:1999tk}. In addition, the $U(1)$ charge of a black hole in the bulk
	corresponds to $R$-charge in the Yang-Mills theory at the boundary
	\cite{Chamblin:1999tk,Behrndt:1998jd,Behrndt:1998ns}. 
In the spirit of this correspondence, the Hawking-Page phase transition between the stable large Schwarzschild black hole and thermal gas in the AdS space \cite{Hawking:1982dh} can also be interpreted as a confinement/deconfinement phase transition in the dual strongly coupled gauge theory \cite{Witten:1998zw}.
	
	 Therefore,
	by defining dimensionless variables,  $x\equiv r_h/L$ and
	$y\equiv q/L^{d-3}$, and inserting $G_{d}$ from Eq. (\ref{gd}) 
	 into Eq. (\ref{mass-entropy}), the internal energy and entropy of the boundary CFT can be written as
	\begin{equation}\label{mass-entropy1}
	U=M=\frac{ (d-2) \hbar\gamma   N^{\frac{d-1}{2}}}{4 L}  \left(y^2
	x^{3-d}+x^{d-3}+x^{d-1}\right), \hspace{0.5cm} 	S
	=\pi  \gamma  N^{\frac{d-1}{2}} x^{d-2}
	\end{equation}
	
The internal energy of the CFT is a quantum mechanical quantity that  vanishes as
	$\hbar \to 0$ by keeping $x$, $y$ fixed, whereas the mass on the AdS side is
	classical.
	It is also convenient to define the dimensionless charge corresponding to R-charge in the  boundary field theory  
	\begin{equation}\label{Qformula}
	\tilde Q=\frac{Q  L}{4 \pi \hbar G_d}= \sqrt{\frac{(d-2)(d-3)}{2}} \gamma N^{\frac{d-1}{2}} y
	\end{equation}
	using Eq. (\ref{gd}). 
	In accord with the first law of thermodynamics, $dU=T dS+\Phi d \tilde Q-P d\mathcal{V}$,  Eq. (\ref{mass-entropy1}) implies that other thermodynamic variables such as temperature, chemical potential, and pressure are  
		\begin{eqnarray}
	T&=&\Big(\frac{\partial U}{\partial S}\Big)_{\tilde Q,\mathcal{V}}=\frac{\{U,\tilde Q,\mathcal{V}\}_{x,y,L}}{\{S,\tilde Q,\mathcal{V}\}_{x,y,L}}=\frac{\hbar}{4 \pi L}
	\left((d-1)x+\frac{d-3}{x}-(d-3)\frac{y^2}{x^{2d-5}}\right),\label{Tformula}\\
	\Phi&=&\Big(\frac{\partial U}{\partial \tilde Q}\Big)_{S,\mathcal{V}}=\frac{\{U,S,\mathcal{V}\}_{x,y,L}}{\{\tilde Q,S,\mathcal{V}\}_{x,y,L}}=\frac{(d-2) \hbar
		y x^{3-d}}{ \sqrt{2(d-3) (d-2)} L},\label{phiformula}\\
	P&=&-\Big(\frac{\partial U}{\partial \mathcal{V}}\Big)_{S,\tilde Q}=-\frac{\{U,S,\tilde Q\}_{x,y,L}}{\{\mathcal{V},S,\tilde Q\}_{x,y,L}}=\frac{  \hbar  \gamma N^{\frac{d-1}{2}} \left(y^2 x^{3-d}+x^{d-3}+x^{d-1}\right)}{4 L^{d-1} \Omega_{d-2}}
	\end{eqnarray}
	In the above equations, we have used Nambu bracket notation \cite{Mansoori:2014oia, HosseiniMansoori:2019jcs}. In the context of gauge/gravity duality, $T$ can be interpreted as the temperature of the quark-gluon plasma \cite{Witten:1998qj,Chamblin:1999tk}. The thermodynamically-conjugate variable to the volume of the CFT, i.e. $\mathcal{V}$, is the pressure, which leads to the $P-\mathcal{V}$ criticality of a boundary CFT \cite{Dolan:2016jjc}. Moreover, $\Phi$ corresponds to the chemical potential associated with R 	current in the dual supersymmetric Yang-Mills theory.
	
\begin{figure}[h]
		\centering
		\includegraphics[width=7cm,height=5.2cm,angle=0]{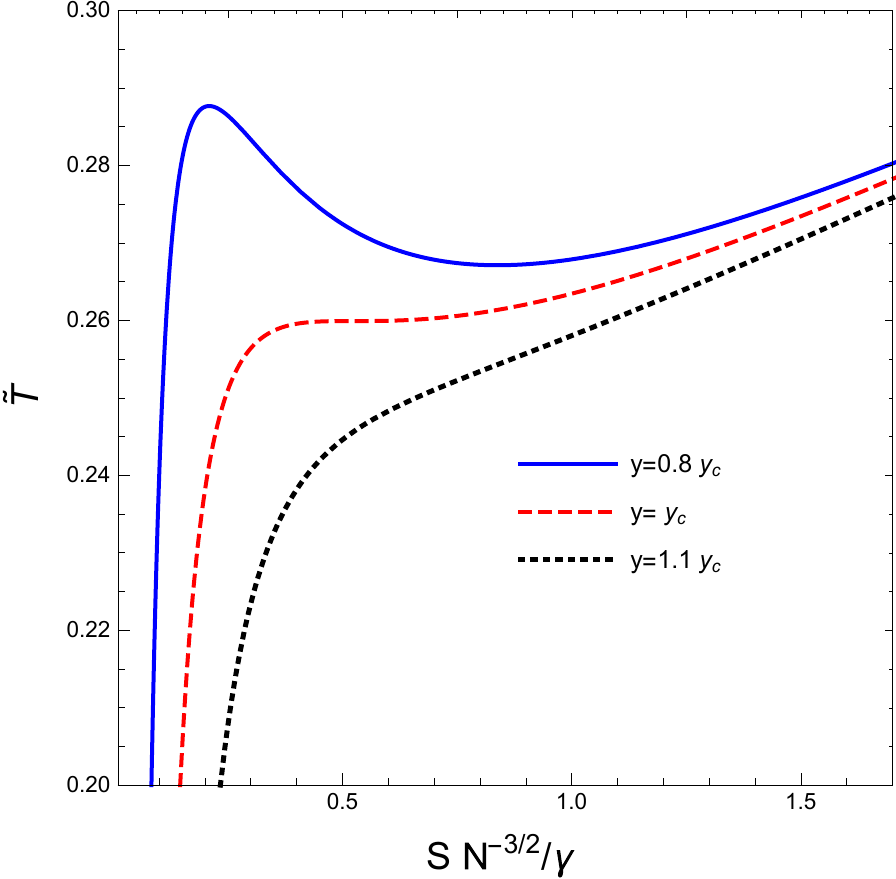}
		\includegraphics[width=7cm,height=5.2cm,angle=0]{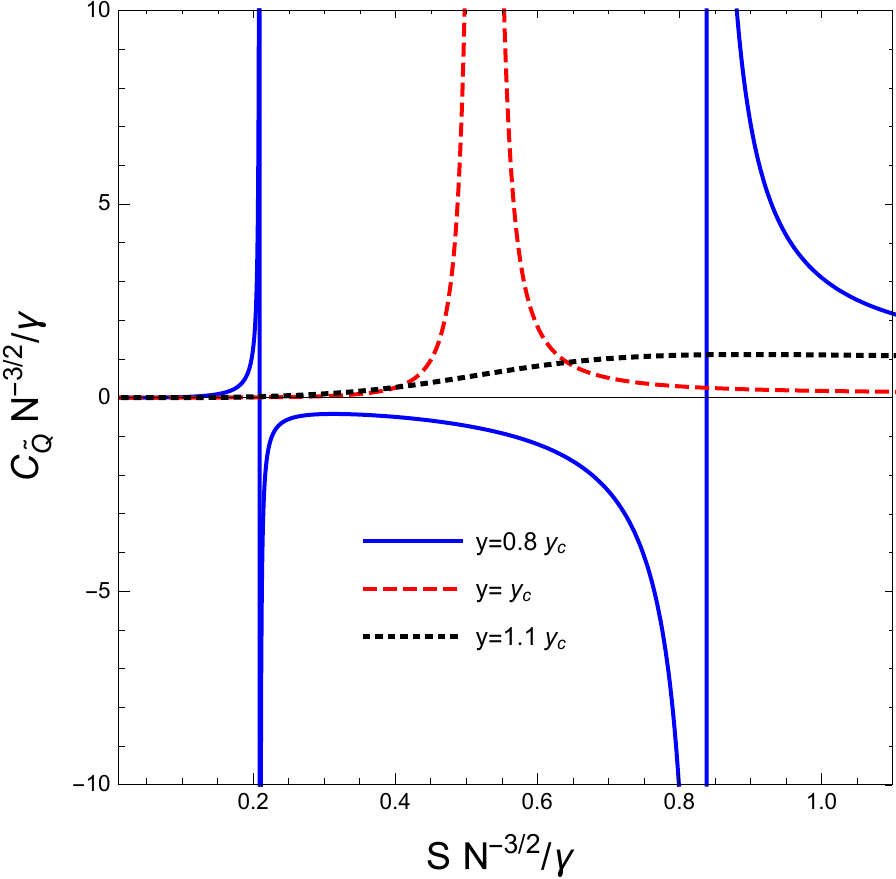}
		\caption{\textbf{Left}: Diagram of the temperature $\tilde T$ as a function of the entropy $S N^{-3/2}/\gamma$. \textbf{Right}:  Diagram of the heat capacity $C_{\tilde Q} /\gamma N^{3/2} (\times 10^{-1})$  versus entropy $S N^{-3/2}/\gamma$. Both plots have been plotted for a triplet $(D; d; k)=(11; 4; 3)$ or 3D CFT while $\hbar=1$.} \label{TS1}
	\end{figure}

	In order to study the phase structure of the boundary CFT, let us first consider 
	 isocharge in a $T-S$  phase diagram, as shown in the left diagram of Fig. \ref{TS1}. We see that
	there is a critical charge   $y_{c}$, at which there is an inflection point
	\begin{equation}
	\Big(\frac{\partial T}{\partial S}\Big)_{\tilde Q,\mathcal{V}}= \Big(\frac{\partial^2
		T}{\partial^2 S}\Big)_{\tilde Q,\mathcal{V}}=0
	\end{equation}
yielding
	\begin{equation}
	x_{c}=\frac{d-3}{\sqrt{(d-1) (d-2)}} \hspace{0.5cm} y_{c}=\frac{(d-3)^{d-3}}{(d-2)^{\frac{d-2}{2}} (d-1)^{\frac{d-3}{2} }\sqrt{2 d-5}}.
	\end{equation}
The critical temperature, pressure, chemical potential, and charge are therefore 
	\begin{eqnarray}\label{criticalpoint1}
	&&T_{c}=\frac{(d-3) \sqrt{(d-1) (d-2)} \hbar}{\pi  (2 d-5) L}, \hspace{0.5 cm}  P_{c}= \frac{\gamma  (d-3)^{d-3}  ((d-5) d+7) \hbar \gamma N^{\frac{d-1}{2}}}{(d-2)^{\frac{d-3}{2} } (d-1)^{\frac{d-1}{2} } (2 d-5) \Omega_{d-2}  L^{d-1}} \nonumber \\
	&&  \Phi_{c}=\frac{\hbar}{ \sqrt{2 (d-3) (2 d-5)} L} \hspace{0.5cm}  \tilde Q_{c}=\frac{ (d-3)^{\frac{2d- 5}{2}} \gamma N^{\frac{d-1}{2}}}{(d-2)^{\frac{d-3}{2}} (d-1)^{\frac{d-3}{2}} \sqrt{4 d-10}}.
	\end{eqnarray}
	This implies that the relevant (dimensionless) parameter that determines the phase transition is $T\mathcal{V}^{\frac{1}{d-2}}$ (or $TL$) and $\mathcal{V}$ and $T$ are not fixed separately \cite{Chamblin:1999tk,Dolan:2016jjc}. In the other words, since the boundary field theory is conformal, $\mathcal{V}$ and $T$ are not independent. It is therefore convenient to consider charge $\tilde Q$ ($y$) rather than $T$ as the control parameter  in discussing the thermodynamics of the system. This latter view is more in keeping with the notion of the quantum phase transition rather than the thermal phase transition \cite{Dolan:2016jjc}.
	
	Furthermore the compressibility factor $Z$ is  
	\begin{equation}
	Z=\frac{\mathcal{V}_{c} P_{c}}{T_{c}}=\frac{\pi   (d-3)^{d-4} ((d-5) d+7) \gamma N^{\frac{d-1}{2}}}{(d-2)^{\frac{d-2}{2}} (d-1)^{d/2}} 
	\end{equation}
which has a constant value for each dimension, as we keep $N$ fixed. 
	For future convenience, we work with dimensionless parameters such as $\tilde{T}= T L$ and $\tilde{\Phi}=\Phi L$ in the rest of the paper.
	
	At the critical point, the phase transition becomes second-order and the heat capacity diverges. The heat capacity at constant charge for the boundary theory is
	\begin{eqnarray}
	C_{\tilde Q} &=&\tilde{T}\Big(\frac{\partial S}{\partial \tilde{T}}\Big)_{ \tilde Q} =\tilde T\frac{\{S,\tilde Q \}_{x,y}}{\{\tilde{T} ,\tilde Q \}_{x,y}} \nonumber \\
	&=& \frac{\pi  (d-2) \gamma N^{\frac{d-1}{2}} x^{d-2} \left(x^{2 d} \left((d-1) x^2+d-3\right)-(d-3) x^6 y^2\right)}{(2 d-5) (d-3) x^6 y^2+\left((d-1) x^2-d+3\right) x^{2 d}}.
	\end{eqnarray} 
	Clearly, $C_{\tilde Q}$ diverges at the critical point in the CFT, whereas heat capacity at fixed charge and volume for the black hole in the bulk is finite at the critical point \cite{Kubiznak:2012wp}.
	The behavior of the heat capacity $C_{\tilde Q}$ with respect to $S$ is illustrated in the right diagram in Fig. \ref{TS1}.
	For small charge $y<y_{c}$, we clearly see that there are two divergent points.
	The first and the third regions at small and large values of the entropy have positive heat capacity 
	whereas the second region in the middle has negative heat capacity. Positive/negative values of the heat capacity imply
	thermodynamic stability/instability.  By increasing $y$, the region of negative heat capacity shrinks, and as  $y\to y_{c}$, the negative region disappears. There then exists only one divergent point at which the heat capacity goes to positive infinity. 
	Further increasing $y$, we see that the divergent behavior of the heat capacity $C_{\tilde Q}$ completely disappears, shown
	as the black dashed line.
	
	Using the pressure as defined above, the $P-\mathcal{V}$ criticality of a boundary CFT
	was investigated explicitly  \cite{Dolan:2016jjc}. However there are a number of differences between the $P-\mathcal{V}$ behavior of the boundary field theory and that of the black
	hole in the bulk. First, in the boundary theory there is a single phase at temperatures below the critical temperature and
	two phases above it.  
	Second, it is   pressure rather than   volume, that
	jumps across the two-phase regime\footnote{This jump may be viewed as a change in the number of degrees of freedom. Thus, any change in the number of
		effective degrees of freedom can contribute to a change in pressure across the phase transition.}. Finally, as one keeps $\mathcal{V}$ and $\tilde{ Q}$ fixed, the critical exponents are not mean field in spite of the fact that the equation of state of the CFT is  similar to
	the Van der Waals equation of state. 
	However, with pressure as the order parameter  in the $P-\mathcal{V}$ plane, the critical exponents are mean field and the
	phase transition is similar to that for a Van der Waals black hole like in the bulk \cite{Dolan:2016jjc}. However, if $\tilde{\Phi}$ instead of $\tilde{Q}$ is taken as the order
	parameter in the $\tilde \Phi- \tilde Q$ diagram, the critical exponents become mean field.
	
	We have already noted a constraint on the thermodynamic variables that arises because the boundary field  theory is conformal, namely that only the combination $T \mathcal{V}^{\frac{1}{d-2}}$ is relevant to the phase structure, not $\mathcal{V}$ and $T$	separately. Consequently, in order understand phase behaviour in terms of $P$ and $\mathcal{V}$,  one needs to work with the dimensionless variables  $P/N^{\frac{d-1}{2}}T^{d-1}$ and $\mathcal{V} T^{d-2}$, and use  charge rather than the temperature as the control
	parameter \cite{Dolan:2016jjc}.  We shall instead extract mean field critical exponents by considering 
phase behaviour in the	$\tilde \Phi- \tilde Q$ plane, since these variables  do not scale directly with the temperature parameter. 	\begin{figure}\centering
		
		\includegraphics[width=7cm,height=5.2cm,angle=0]{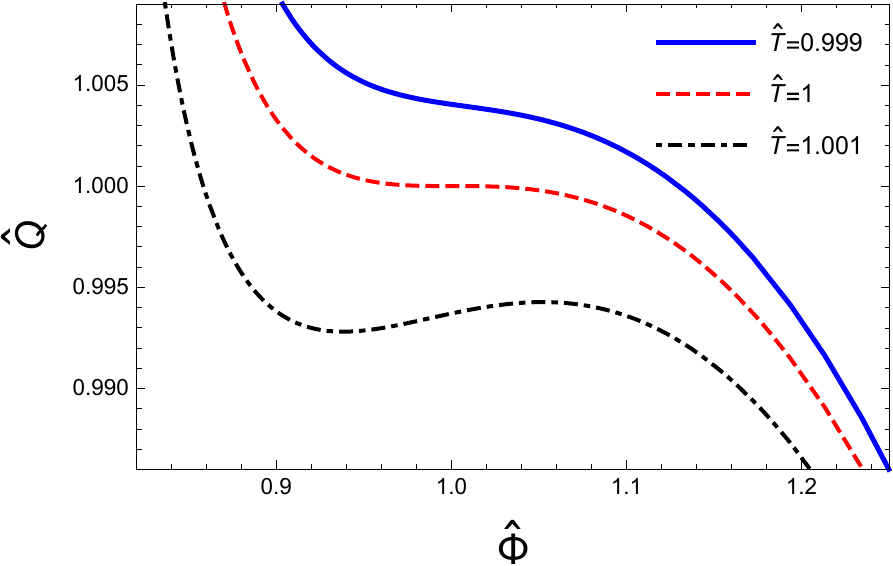}
		\includegraphics[width=7cm,height=5.2cm,angle=0]{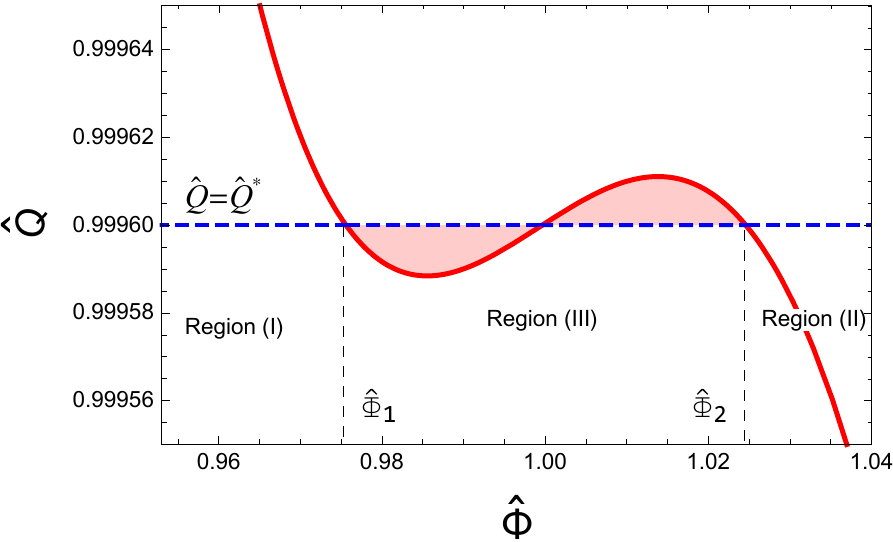}
		\caption{ \textbf{Left}: Behavior of the $\hat{Q}$ as a function of $\hat{\Phi}$ for different temperatures $\hat{T}=0.999$, $\hat{T}=1$ and $\hat{T}=1.001$. Maxwell equal area rule can be applied for $\hat{T}>1$. \textbf{Right}: Behavior of the $\hat{Q}$ as a function of $\hat{\Phi}$ for $\hat{T}=1.001$ and Maxwell equal area rule has been shown. Both sides have been obtained from a triplet $(D; d; k)=(11; 4; 3)$ or 3D CFT with $\hbar=1$. } \label{maxwell}
	\end{figure}

	By eliminating $x$ and $y$ in favour of $\hat{Q}=\frac{Q}{Q_c}=\frac{\tilde{Q}}{\tilde{Q}_c}$ and $\hat{\Phi}=\frac{ \Phi}{\Phi_c}=\frac{\tilde \Phi}{\tilde{\Phi}_c}$,  upon inserting   Eqs. (\ref{Qformula}) and (\ref{phiformula}) into Eq. (\ref{Tformula}) we obtain
	\begin{eqnarray}\label{Tem}
	\tilde{T}=&&\frac{(d-3)^{\frac{1}{3-d}}  \hbar^{\frac{d-2}{3-d}}\gamma ^{\frac{1}{3-d}} N^{\frac{d-2}{d-3}} \tilde Q^{\frac{1}{d-3}} \tilde \Phi ^{\frac{1}{3-d}}}{4 \pi  (d-2)} \nonumber \\
	&&(d-2) (d-1) h^{\frac{d-2}{d-3}} N^{\frac{d}{3-d}}+(d-3)^{\frac{d-1}{d-3}} \gamma ^{\frac{2}{d-3}} N^{\frac{1}{3-d}} \tilde Q^{\frac{2}{3-d}} \tilde \Phi ^{\frac{2}{d-3}} \left((d-2) \hbar^2-2 (d-3) \tilde \Phi ^2\right) \nonumber \\
	\end{eqnarray}
for  the equation of state.  We depict the resultant phase behaviour in  Fig. \ref{maxwell}, whose left diagram 
illustrates isotherm curves in the $\hat{\Phi}-\hat{Q}$ plane. The behaviour is reminiscent of a 
reverse Van der
	Waals transition\footnote{ It is worthwhile noted that there is the Maxwell construction for $\hat T<1$ in charged AdS black holes.} \cite{kubizvnak2017black,Frassino:2014pha}, with an inflection point at $\hat{T} = 1$ and two distinct phases at larger values of $\hat{T}$.
	 As a consequence, the critical point derived in Eq. (\ref{criticalpoint1}) is also obtained from 
	\begin{equation}
	\Big(\frac{\partial  \hat Q}{\partial \hat \Phi}\Big)_{\hat T}= \Big(\frac{\partial^2
		\hat Q	}{\partial^2 \hat \Phi}\Big)_{\hat T}=0.
	\end{equation}

From the above similarities,  formally identifying the variables $( \hat{Q}, \hat{\Phi})$
	of the boundary field theory with $(P-V)$ of a van der Waals liquid gas system,  the phase structure of the boundary field theory mimics qualitatively certain 
	remarkable properties to that of a van der Waals liquid gas system. Note that it is a reverse VdW transition, with one
	phase at cold temperatures and two phases at hot temperatures.
An order parameter in the boundary field theory that
 measures the phase change across the critical point can  be defined in terms of the Maxwell equal-area law as shown in 
 the right diagram of Fig. \ref{maxwell}.  Analogous to a Van der Waals system, we therefore define
	\begin{equation}
	\eta=|\hat \Phi_{2}-\hat \Phi_{1}|
	\end{equation}
	as the order parameter to characterize the phase change of the boundary CFT near the critical point.  For $\hat{T}>1$ there is a two-phase regime in the boundary system. To describe this phase transition one can replace the oscillating part of the isotherm in Fig. \ref{maxwell} by an isocharge from the equal area relation
	\begin{equation}\label{maxwellrelation}
	\int_{\hat{\Phi}_1}^{\hat{\Phi}_2} \hat{Q} d\hat{\Phi}= \hat{Q}^*(\hat{\Phi}_2)-\hat{Q}^*(\hat{\Phi}_1)
	\end{equation} 
where $\hat{Q}$ is regarded as a function of $(\hat{T},\hat{\Phi})$ from  \eqref{Tem}, i.e., 
	\begin{eqnarray}
	\hat{Q}&=& \frac{\left(4 (d-2) \hat T \hat \Phi ^{\frac{5}{d-3}}-\frac{2 A}{\sqrt{d-3}}\right)^{d-3}}{(4 d-10)^{d-3} \hat \Phi ^4} ,\\
	A&=& \sqrt{(d-2) \left(4 (d-5) d \left(\hat T^2-1\right)+24 \hat T^2-25\right) \hat \Phi ^{\frac{10}{d-3}}+(2 d-5) \hat \Phi ^{\frac{2 (d+2)}{d-3}}}
	\end{eqnarray}
We define
	\begin{equation}
	t=\hat{T}-1, \hspace{0.5cm} \phi=\hat{\Phi}-1.
	\end{equation}
in order to characterize critical exponents  describing the behaviour of physical
	quantities near the critical point.  
	
\begin{figure}\centering
		\includegraphics[width=7cm,height=5.2cm,angle=0]{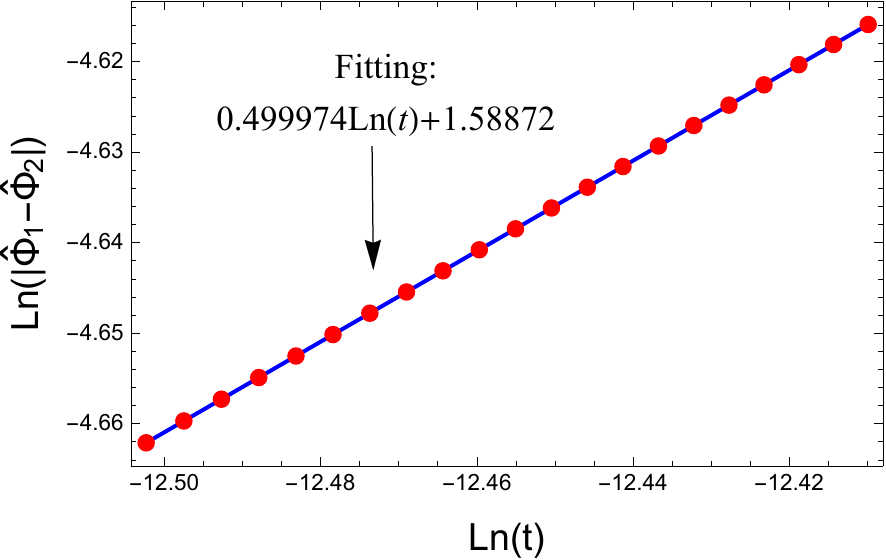}
		\caption{Diagram of $ln(|\hat{\Phi}_1-\hat{\Phi}_2|)$ versus $ln(t)$. The fitted straight line for the data points (red dot) is given by $\Delta \hat{\Phi}= b(t^\beta)$ with $b =1.58872$ and $\beta = 0.0.499974$ for a triplet $(D; d; k)=(11; 4; 3)$ or 3D CFT while $\hbar=1$. \label{Delta_Phi}}
	\end{figure}
	
	The critical exponent $\beta$, which characterizes the behavior of the chemical potential difference $\eta$, along the $\hat{Q}-\hat T$ coexistence curve, is defined as
	\begin{equation}\label{delta_phi}
	\eta=|\hat{\Phi}_1-\hat{\Phi}_2| =  b t^\beta \hspace{0.5cm} \text{for} \hspace{0.5cm} t>0
	\end{equation} 	
	whose behaviour we illustrate in 	
	Fig. \ref{Delta_Phi}, plotting $\ln(|\hat \Phi_{1}-\hat \Phi_{2}|)$ as a function of $\ln(t)$. The fitted straight line for the
	data points is generated by solving Eq. \eqref{maxwellrelation}numerically  for $t>10^{-4}$. We find  $b =  1.58872$ and $\beta=0.499974$. 
	
	In the reduced parameter space, the isothermal compressibility $\kappa_{\tilde T}$,  and the adiabatic compressibility $\kappa_{S}$, are respectively given by 
	\begin{eqnarray}\label{KT}
	\kappa_{\tilde T}&=&-\frac{1}{\tilde \Phi} \Big(\frac{\partial \tilde \Phi}{\partial \tilde Q}\Big)_{\tilde T}=-\frac{1}{\hat Q_{c}\hat \Phi} \Big(\frac{\partial \hat \Phi}{\partial \hat  Q}\Big)_{\hat T}\\ \nonumber \kappa_{S}&=&-\frac{1}{\tilde \Phi} \Big(\frac{\partial \tilde \Phi}{\partial \tilde Q}\Big)_{S}=-\frac{1}{\hat Q_{c} \hat \Phi} \Big(\frac{\partial \hat \Phi}{\partial \hat  Q}\Big)_{\hat S} 
	\end{eqnarray}
	By defining the heat capacity at constant chemical potential, $C_{\tilde \Phi}=\tilde T\Big(\partial S/\partial \tilde T\Big)_{\tilde \Phi}$, we find a fascinating relation 
	\begin{equation}
	\kappa_{\tilde T} C_{\tilde \Phi} \Big(\kappa_{S}C_{\tilde Q}\Big)^{-1}=1.
	\end{equation}
between heat capacities and compressibilities\footnote{This can be shown using the bracket notation. We have
		\begin{equation}
		\kappa_{\tilde T} C_{\tilde \Phi} \Big(\kappa_{S}C_{\tilde Q}\Big)^{-1}=\Big(\frac{\partial \tilde \Phi}{\partial \tilde Q}\Big)_{\tilde T} \Big(\frac{\partial S}{\partial \tilde T}\Big)_{\tilde \Phi} \Big(\frac{\partial  \tilde Q}{\partial \tilde \Phi}\Big)_{S} \Big(\frac{\partial \tilde T}{\partial S}\Big)_{\tilde Q}=\frac{\{\tilde \Phi,\tilde T\}}{\{\tilde Q,\tilde T\}}\frac{\{S,\tilde \Phi\}}{\{\tilde T,\tilde \Phi\}}\frac{\{\tilde Q,S\}}{\{\tilde \Phi,S\}}\frac{\{\tilde T,\tilde Q\}}{\{S,\tilde Q\}}=1
		\end{equation}
		Note that all brackets are written in the coordinates $(\tilde T,\tilde \Phi)$.},	
	Generally, near the critical point heat capacity and compressibility exhibit the critical behaviour 
	\begin{equation}\label{criticalck}
	C_{I} \approx
	\begin{cases}
	t^{-\alpha_{I}} & t>0  \\
	(-t)^{-\alpha_{I}'} & t<0  \\
	\end{cases}\hspace{0.5cm}\text{and}\hspace{0.5cm}\kappa_{I} \approx
	\begin{cases}
	t^{-\gamma_{I}} & t>0  \\
	(-t)^{-\gamma_{I}'} & t<0  \\
	\end{cases}
	\end{equation}
along the
	$\hat{Q}-\hat{\Phi}$ coexistence curve and isochemical potential $\hat{\Phi}=1$ line, respectively.
		
Along the isochemical potential curve  $\hat{\Phi}=1$,  substituting $\hat T = 1+(-t)$ into the  heat capacity and compressibility functions and then expanding them to  lowest order in $(-t)$ yields
	\begin{eqnarray}\label{critcalc}
	C_{\tilde Q}&\approx& \frac{\pi  (d-3)^{d-2} \gamma N^{\frac{d-1}{2}}}{(d-2)^{\frac{d-4}{2}} (d-1)^{\frac{d-2}{2}} (2 d-5) (-t)} \hspace{0.25cm} \\
	C^{c}_{\tilde \Phi} &=&-\frac{2 \pi  (d-3)^{d-2} \gamma N^{\frac{d-1}{2}}}{(d-2)^{\frac{d-6}{2}} (d-1)^{\frac{d-2}{2}}}\\
	\kappa_{\tilde T} &\approx& \frac{(d-2)^{\frac{d-5}{2}} (d-1)^{\frac{d-3}{2}}}{\sqrt{4 d-10} (d-3)^{\frac{2 d-5}{2} } \gamma  N^{\frac{d-1}{2}} (-t)} \hspace{0.25cm}\\
	\kappa^{c}_{S} &=&-\frac{((d-2) (d-1))^{\frac{d-3}{2}} \sqrt{4 d-10}}{(d-3)^{\frac{2d-5}{2}} \gamma N^{\frac{d-1}{2}}}
	\end{eqnarray}
	which determine $\alpha'_{\tilde Q}=\gamma'_{\tilde T}=1$ and $\alpha'_{\tilde \Phi}=\gamma'_{S}=0$ for $t<0$. As indicated  above, the adiabatic compressibility $\kappa_{S}$ and the heat capacity $C_{\tilde \Phi}$ are negative at the critical point, reflecting  the instability of the system. Similar instabilities have been observed for charged AdS black holes \cite{Chamblin:1999hg} and  rotating AdS black holes  with constant angular momentum \cite{Caldarelli:1999xj,Dolan:2014lea}.

\begin{figure}\centering
		\includegraphics[width=7cm,height=5.2cm,angle=0]{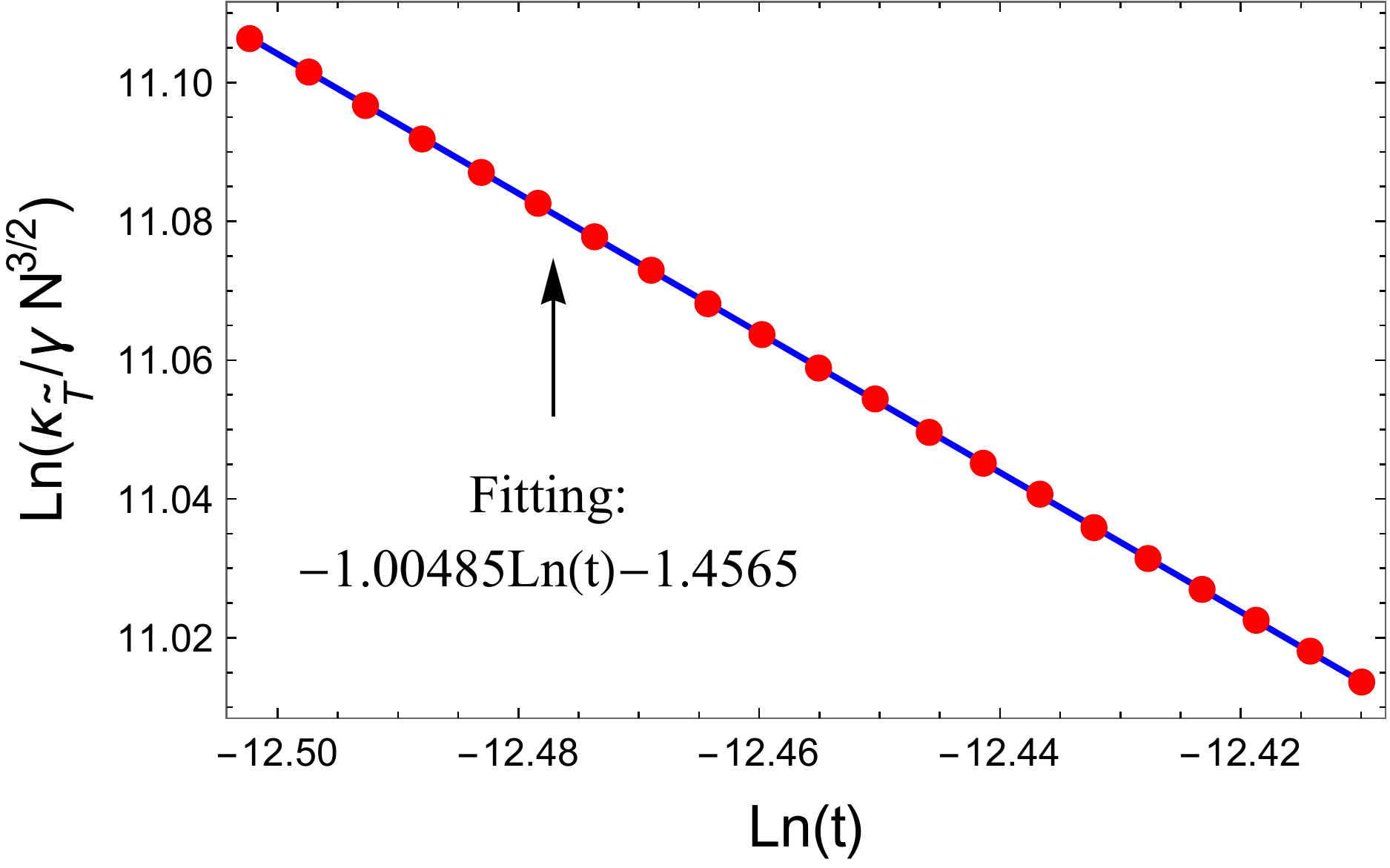}
		\includegraphics[width=7cm,height=5.2cm,angle=0]{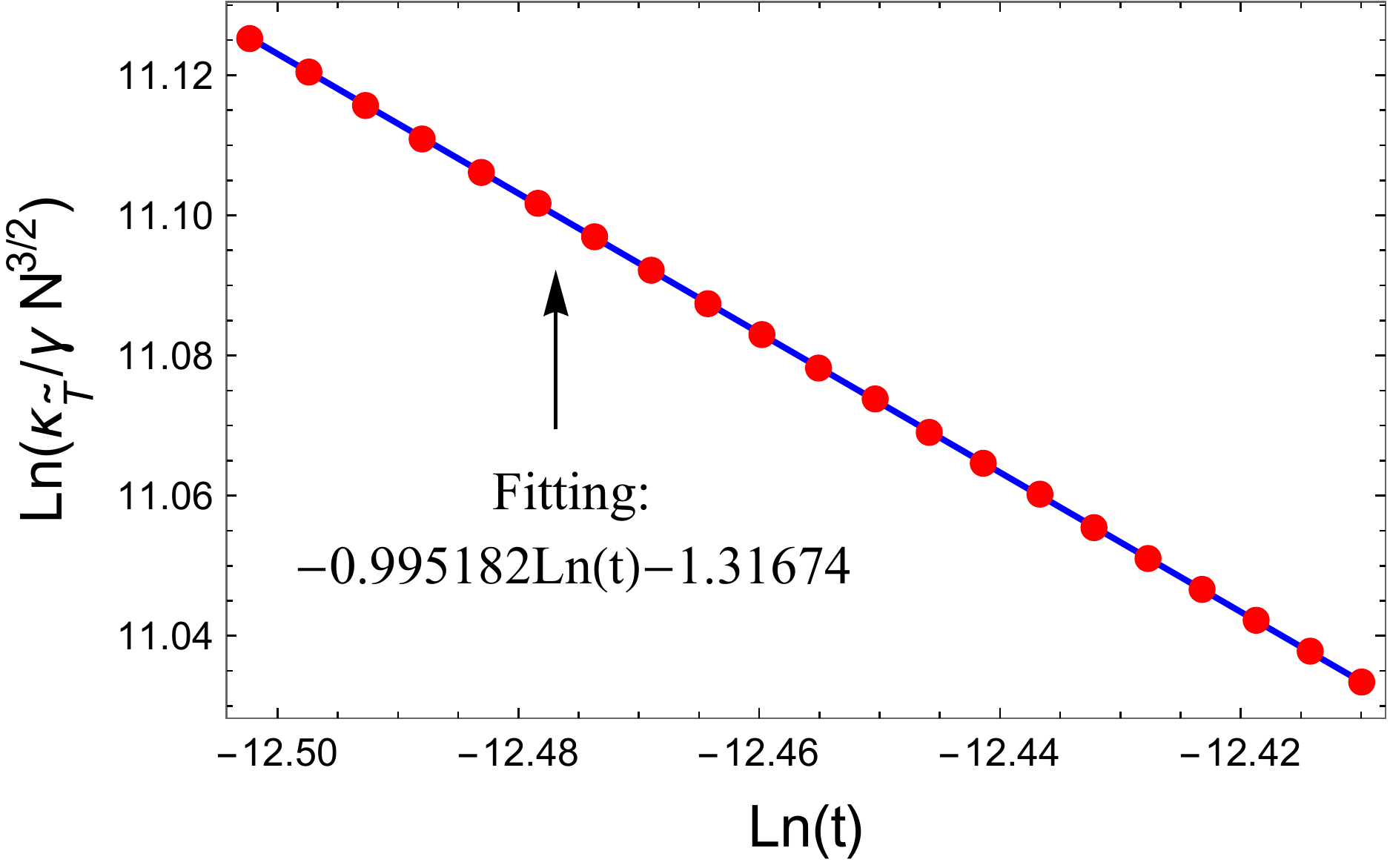}
		\caption{Diagram of $ln(\kappa_{\tilde T}\gamma N^{3/2})$ versus $ln(t)$ for a triplet $(D; d; k)=(11; 4; 3)$ or 3D CFT while $\hbar=1$. The fitted straight line for the data points (red dot) is given by $\kappa_{\tilde T} \gamma N^{3/2}= r (t^{-\gamma_{\tilde{ T}}})$. \textbf{Left}, for region (I), $r =-1.4565$ and $\gamma_{\tilde{ T}} = 1.00485$. \textbf{Right}, for region (II) $r =-1.31674$ and $\gamma_{\tilde{ T}} = 0.995182$.}\label{Fig_KT}
	\end{figure}

We illustrate in 	Fig. \ref{Fig_KT} numerical plots of isothermal compersibility $\kappa_{\tilde T}$ versus $t$ for
regions (I) (left) and (II) (right) in Fig.~\ref{maxwell} along the 	$\hat{Q}-\hat{\Phi}$ coexistence curve for the  triplet $(11;4;3)$.  We obtain the predicted asymptotic critical behavior $\kappa_{\tilde T} \approx (t)^{\gamma_{\tilde T}}$ with exponent $\gamma_{\tilde T}=1$. The coexistence saturated regions (I) and (II)  may   respectively be interpreted as a gas of particles in quark-gluon plasma phase and hadrons in the confinement phase. Similar results are calculated for other triplets, which we collect in table \ref{tab1}.
Note  that if pressure is taken as the order parameter (whose value jumps across the line of first order phase transitions
in the $P-\mathcal{V}$ plane) and $\tilde Q$ as the control variable the exponents are always mean
	field in the large $N$ limit \cite{Dolan:2016jjc}.
	
	\begin{table}[h]
		\begin{center}
			\begin{tabular}{c||c c c c c}
				\hline\hline
				Critical exponent & $\beta$ & $\alpha_{\tilde{Q}}$ & $\gamma_{\tilde T}$ \\ \hline
				(11;4;3); 3D CFT
				&  0.49997  & 0.999892 &    1.00002  \\\hline
				(10;5;0); 4D CFT
				& 0.49996 &  0.999919 & 1.00012  \\\hline
				(11;7;-3); 6D CFT
				& 0.49999 & 1.00003 &  1.00055  \\ \hline\hline
			\end{tabular}
			\caption{ Critical exponent values of $\eta$, $C_{\hat Q}/\gamma N^{\frac{d-1}{2}}$ and $\kappa_{\tilde T}\gamma N^{\frac{d-1}{2}}$ for $t>0$. These values have been averaged between the values obtained from region (I) and (II).}\label{tab1}
		\end{center}
	\end{table}
	

	\section{Criticality of Thermodynamic curvatures}\label{TCG}
	
 Another perspective on the thermodynamic properties of a system that has grown in interest in recent years entails  geometric methods \cite{reff8,reff9,reffff9,reff12,reff10,reff4,reff5}. A metric is constructed on the thermodynamic phase space,
and its  Riemann scalar curvature  is found to reveal some information about black hole phase transitions and criticality. 
It also provides us with  powerful tools   for understanding microstructures of thermodynamic systems \cite{reff9,Wei:2015iwa,Wei:2019uqg,Wei:2019yvs}.

However, in some contradictory examples \cite{reff10,Sarkar:2006tg} the well-known Ruppeiner geometry, which was defined as a Hessian matrix of the entropy, was not successful in achieving a one-to-one correspondence between phase transition points and singularities of the corresponding scalar curvature.  To address this issue, a new formulation of  the geometry 
of thermodynamic phase space 
was developed, based on considerations about  thermodynamic potentials related to  mass (instead of the entropy) by Legendre transformations  \cite{Mansoori:2013pna,HosseiniMansoori:2019jcs,Mansoori:2016jer}. This new formalism of thermodynamic geometry (NTG) equips us with a one-to-one despondence between critical points where heat capacity diverges and curvature singularities.  The extrinsic curvature of a certain kind of hypersurface immersed in thermodynamic space constructed by NTG metric not only is singular at the phase transition point, but also has the same sign as the heat capacity around the critical point \cite{Mansoori:2016jer}.  However  it fails to explain the thermal stability of a thermodynamic system around a critical point.

\begin{figure}\centering
		\includegraphics[width=7cm,height=5.2cm,angle=0]{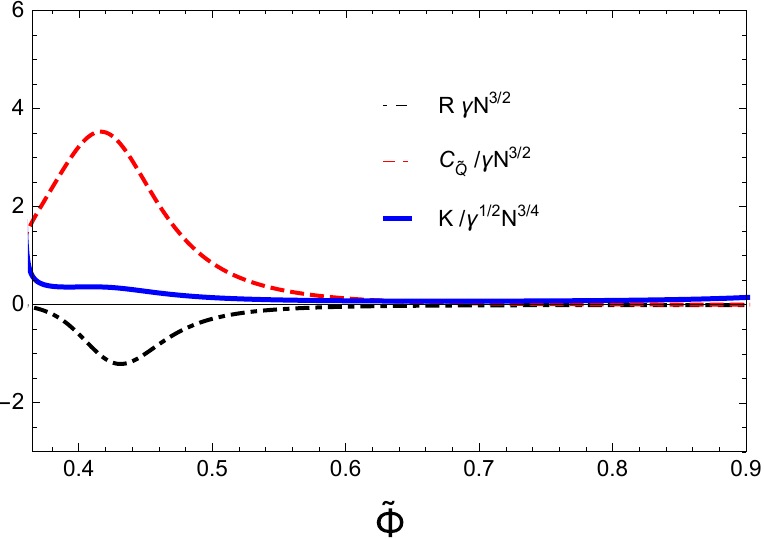}
		\includegraphics[width=7cm,height=5.2cm,angle=0]{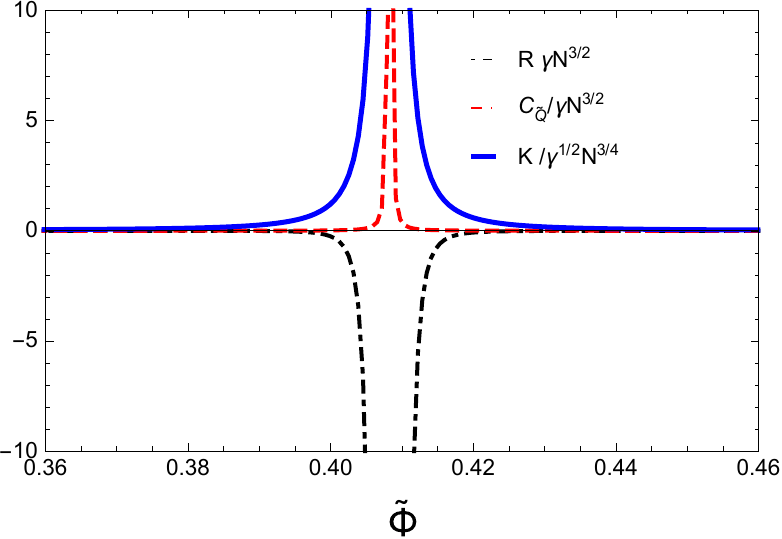}
		\includegraphics[width=7cm,height=5.2cm,angle=0]{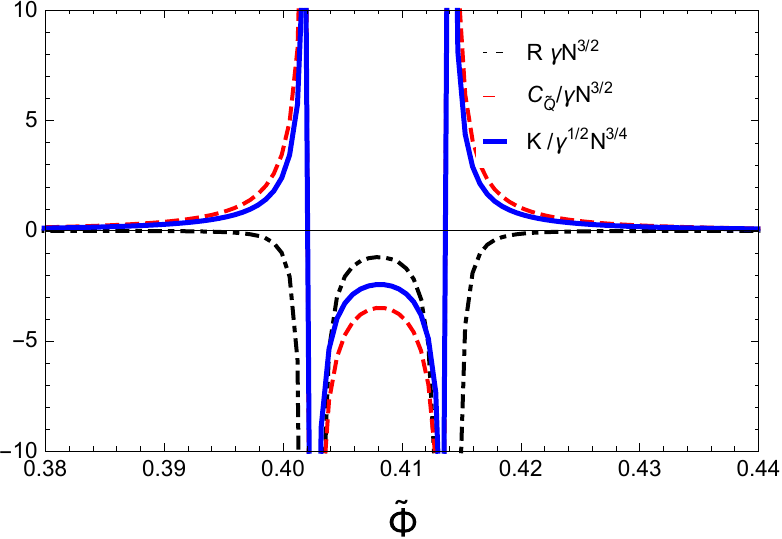}
		\caption{ Behavior of  Ricci scalar, heat capacity $C_{\tilde Q}/\gamma N^{\frac{3}{2}}$ and extrinsic curvatue with respect to $\tilde \Phi$, \textbf{Top(Left)}: for $t<0$, $R\gamma N^{\frac{3}{2}}(\times 10^{-3})$, $C_{\tilde Q}/\gamma N^{\frac{3}{2}}(\times 10^{-1})$  and $ K \gamma^\frac{1}{2} N^{\frac{3}{4}} (\times 10^{-1})$ have been plotted. \textbf{Top(Right)}: for $t=0$,  $R(\times 10^{-7})\gamma N^{\frac{3}{2}}$, $C_{\tilde Q}/\gamma N^{\frac{3}{2}}(\times 10^{-5})$  and $K \gamma^\frac{1}{2} N^{\frac{3}{4}} (\times 10^{-4}) $ have been plotted. \textbf{Bottom}: for $t>0$, $R\gamma N^{\frac{3}{2}}(\times 10^{-7})$, $C_{\tilde Q}/\gamma N^{\frac{3}{2}}(\times 10^{-3})$  and $ K \gamma^\frac{1}{2} N^{\frac{3}{4}}(\times 10^{-3})$ have been depicted. \label{R_CQ_2}}
	\end{figure}

From the second holographic interpretation of black hole chemistry, in which varying
$\Lambda$ in the bulk corresponds to varying the curvature radius
governing the space on which the field theory is defined, we employ  NTG  \cite{HosseiniMansoori:2019jcs} to study the thermodynamics of the boundary CFT, 
	which is defined by
	\begin{equation}\label{Ru1}
	dl_{NTG}^2=\frac{1}{T}\left( \eta_i^{ j} \, \frac{\partial^2\Xi}{\partial X^j
		\partial X^l} \, d X^i d X^l \right)
	\end{equation}
	where $\eta_i^{ j}={\rm diag} (-1,1,...,1)$,  $\Xi$ is the thermodynamic
	potential, and the $X^{i}$ can be intensive or extensive variables
	\cite{HosseiniMansoori:2019jcs}. Interestingly, this formalism is able to explain a one-to-one correspondence between phase transitions and curvature singularities. For example, by choosing $\Xi=\tilde G(\tilde T, \tilde \Phi)$ where $\tilde G=\tilde U-\tilde T S-\tilde \Phi \tilde Q$ is the Gibbs energy and $X^{i}=(\tilde T,\tilde \Phi)$ in NTG metric
	Eq. \eqref{Ru1}, we have
	\begin{equation}\label{CFTmetric_2}
	g^{NTG}_{\tilde G}=\frac{1}{\tilde T} \text{diag}\Big(-\frac{\partial^2 \tilde G}{\partial \tilde T^2 },\frac{\partial^2 \tilde G}{\partial \tilde \Phi^2 }\Big)=C_{\tilde \Phi} \text{diag}\Big(\tilde T^{-2},-\mathcal{F}\Big)=C_{\tilde \Phi} \hat{g}, \hspace{0.25cm} \text{with} \hspace{0.25cm} \mathcal{F}=\frac{\Big(\frac{\partial \tilde Q}{\partial \tilde \Phi}\Big)_{\tilde T}}{\tilde T C_{\tilde \Phi}}
	\end{equation}
	in which we have used the first law  $d \tilde G=-Sd\tilde T-\tilde Qd \tilde \Phi$ for the Gibbs free energy. According to the new metric $\hat{g}$, which is conformally equivalent with the previous one, the Ricci scalar can be
	written as \cite{carroll2019spacetime,HosseiniMansoori:2019jcs}
	\begin{equation}\label{intrinsic}
	R=C_{\tilde \Phi}^{-1}\Big[\hat{R}-\Box \ln C_{\tilde \Phi}\Big] 
	\end{equation}
	where $\Box=\partial^\mu \partial_\mu $ is d'Alembert operator associated with \eqref{CFTmetric_2} and 
	\begin{equation}\label{rdot}
	\hat{R}=\tilde T \frac{\partial \ln(\mathcal F)}{\partial \tilde T}\Big[\frac{1}{2} \tilde T \frac{\partial \ln(\mathcal F)}{\partial \tilde T}-1-\tilde T \frac{\partial}{\partial \tilde T}\Big(\ln \Big(\frac{\partial \mathcal F}{\partial \tilde T}\Big)\Big)\Big]
	\end{equation}
	As illustrated in Fig. \ref{R_CQ_2}, the singularities of the Ricci scalar $R^{NTG}$ occur exactly at phase transitions of the heat capacity $C_{\tilde Q}$ at constant charge    with no other
	additional roots\footnote{The conjugate potential $\bar{\Xi}=\tilde U$, which obeys $\Xi+\bar{\Xi}=2 \tilde{U}-\tilde{ T} S-\tilde{ \Phi} \tilde{Q}$, yields the same result \cite{HosseiniMansoori:2020yfj}.
	}. As a consequence
	of the NTG method, there is a one-to-one correspondence between curvature singularities and phase transitions of the heat capacity. 
	
	Although NTG curvature indicates the location of phase transition points, it is not able to explain  thermodynamic stability of a system. As shown in Fig. \ref{R_CQ_2}, the sign of the curvature does not always correspond to the sign of
	the heat capacity.  However it has been shown  \cite{Mansoori:2016jer} that the
	extrinsic curvature of a certain kind of hypersurface immersed in thermodynamic phase space space contains unexpected information about stability of a system. More precisely, its extrinsic curvature not only diverges at phase transition points, but also everywhere has the same sign as the heat capacity.
	
To find this hypersurface, we 	
	consider $\tilde T$ constant hypersufaces in the thermodynamic manifold
	constructed from the metric elements in \eqref{CFTmetric_2}. For such a hypersurface, the unit normal vector is defined as $n_{i}=(1,0)$ and $n^{i}=g^{ij} n_{j}=(\sqrt{|C_{\tilde \Phi}|}/\tilde T,0)$
	. Therefore, the extrinsic curvature can be
	defined by $K=\nabla_{i}n^{i}$ which yields
	\begin{equation}\label{extrinsic}
	K= \frac{\tilde T}{2\sqrt{|C_{\tilde \Phi}|}} \frac{\partial}{\partial \tilde T}\Big(\ln \Big(|C_{\tilde \Phi}|\mathcal{F}\Big)\Big)
	\end{equation}
Fig. \ref{R_CQ_2} shows that  the
	extrinsic curvature of the constant  $\tilde T$ hypersurface has everywhere the same sign as the heat capacity, in contrast to the scalar curvature.

	Let us now examine the intrinsic and extrinsic curvatures obtained in Eqs. \eqref{intrinsic} and \eqref{extrinsic} respectively as   the critical point is approached along the isochemical potential line $\hat{\Phi}=1$ (or $\tilde{\Phi}=\tilde{\Phi}_{c}$). 
Near the critical point we expand   $\hat{Q}=\sum_{i,j}a_{ij}t^{i}\phi^{j}$, where $a_{00}=1$. Since the respective critical behaviour of $\hat{\Phi}$ along the coexistence and isochemical potential lines is $\hat{\Phi}=1$ and $\hat \Phi \approx t^{\beta}$ (with $\beta>1$), it is sufficient to consider the expansion of $\hat Q$ to linear order in $t$.  Therefore, it is easy to see that the last derivative term in Eq. \eqref{rdot} can be completely ignored in comparison with other terms near the critical point.       
	Taking
	advantage of Eqs. \eqref{KT} and \eqref{criticalck} we obtain 
	\begin{equation}
	\mathcal{F}=\frac{\Big(\frac{\partial \tilde Q}{\partial \Phi}\Big)_{\tilde T}}{\tilde T C_{\tilde \Phi}}=-(\tilde T \tilde \Phi C_{\tilde \Phi} \kappa_{\tilde T})^{-1}\approx -\frac{(1+(-t)) (-t)^{\gamma'_{\tilde T}}}{C^{c}_{\tilde \Phi}}\approx -\frac{(-t)^{\gamma'_{\tilde T}}}{ C^{c}_{\tilde \Phi}}  \hspace{0.5cm} \text{for} \hspace{0.5cm} t<0
	\end{equation}
	Because $C_{\tilde \Phi}$ is finite at the critical point, the last term in Eq. \eqref{intrinsic} vanishes, and so
	\begin{eqnarray}
	R_{N}\equiv R C_{\tilde \Phi}^{c}\approx \hat{R}\approx (1+t') \frac{\partial \ln(\mathcal F)}{\partial t'}\Big[\frac{1}{2} (1+t') \frac{\partial \ln(\mathcal F)}{\partial t'}-1\Big]\approx \frac{{\gamma'_{\tilde T}}^2}{2 t'^{2}}
	\end{eqnarray}
We also find 
	\begin{equation}\label{extrinsic}
	K_{N}=K\sqrt{|C_{\tilde \Phi}^{c}|}=\frac{1+t'}{2} \frac{\partial}{\partial t'}\Big(\ln \Big(\mathcal{F}C_{\tilde \Phi}^{c}\Big)\Big) \approx \frac{\gamma'_{\tilde T}}{2 t'}
	\end{equation}
 near the critical point, where we have replaced $T$ by $t'+1$,  with $t'=-t$, to lowest order in $t'$. 
In summary, we find
	\begin{eqnarray}\label{criticalRNS}
	R_{N} &\equiv& R C_{\tilde \Phi}^{c} \approx \frac{{\gamma'_{\tilde T}}^2}{2 t^{2}} \hspace{0.5cm} \text{for} \hspace{0.5cm} t<0\\
	\nonumber 	K_{N} &\equiv& K\sqrt{|C_{\tilde \Phi}^{c}|} \approx -\frac{\gamma'_{\tilde T}}{2 t} \hspace{1cm} \text{for} \hspace{0.5cm} t<0
	\end{eqnarray}
	
As for VdW AdS black holes \cite{HosseiniMansoori:2020jrx, Wei:2019uqg} we can define  normalized intrinsic and extrinsic curvatures. Since the critical exponent $\gamma'_{\tilde T}=1$ in all cases, we see that $R_{N} t^2 =1/2$ and $K_{N} t=-1/2$ for $t <0$, commensurate with the result \cite{HosseiniMansoori:2020jrx} for d-dimensional charged AdS black holes in the bulk for $t>0$. However, there is a sign difference between our result \eqref{criticalRNS} for $R_{N}$ and that for charged AdS black holes  \cite{HosseiniMansoori:2020jrx}, which comes
	from $C_{\tilde \Phi}<0$ at the critical point for the boundary system\footnote{The same issue has been reported for charge AdS black holes in a cavity \cite{Wang:2019cax}. }. This finding demonstrates that there is   universal behaviour of thermodynamics curvature near the critical point, namely, they do not depend on the details
	of the boundary system.
	
To check  this universality property,  we approach the  critical point along the two-phase regime coexistence line. Using Eqs. \eqref{criticalck} and \eqref{delta_phi}, we find
	, one obtains
	\begin{equation}
	\mathcal{F}=-(\tilde T \tilde \Phi C_{\tilde \Phi} \kappa_{\tilde T})^{-1}\approx -\frac{(1+t) t^{\gamma_{\tilde T}-\beta}}{C^{c}_{\tilde \Phi}}\approx -\frac{t^{\gamma_{\tilde T}-\beta}}{ C^{c}_{\tilde \Phi}}  \hspace{0.5cm} \text{for} \hspace{0.5cm} t>0
	\end{equation}
to lowest order in $t$. From this we straightforwardly find  
	\begin{eqnarray}\label{criticalRNB}
	R_{N}&\equiv& R C_{\tilde \Phi}^{c}\approx (1+t) \frac{\partial \ln(\mathcal F)}{\partial t}\Big[\frac{1}{2} (1+t) \frac{\partial \ln(\mathcal F)}{\partial t}-1\Big]\approx \frac{{(\beta-\gamma_{\tilde T})}^2}{2 t^{2}}\\
	\nonumber 	K_{N}&=&K\sqrt{|C_{\tilde \Phi}^{c}|}=\frac{1+t}{2} \frac{\partial}{\partial t}\Big(\ln \Big(\mathcal{F}C_{\tilde \Phi}^{c}\Big)\Big) \approx \frac{\gamma_{\tilde T}-\beta}{2 t}
	\end{eqnarray}
	According to the data reported in table \ref{tab1},   the values of the critical
	exponents $\beta$ and $\gamma_{\tilde T}$ are essentially the same for all  CFT dimensions;    we can take $\gamma_{\tilde T} \approx 1$ and $\beta\approx 1/2$. From Eqs. \eqref{criticalRNS} and \eqref{criticalRNB}, 
as $t\to0^{\pm}$  we get	
	\begin{eqnarray}\label{finalcritically}
	R_{N} t^2  \approx \bigg\{ \begin{matrix}
	\frac{1}{2} & \text{for} & t<0,\\
	\frac{1}{8} & \text{for} & t>0,
	\end{matrix} \hspace{0.5cm} \text{and} \hspace{0.5cm} K_{N} t  \approx \bigg\{ \begin{matrix}
	-\frac{1}{2} & \text{for} & t<0,\\
	\frac{1}{4 } & \text{for} & t>0.
	\end{matrix}
	\end{eqnarray} 
	
	We emphasize that we approach the critical
	point along different paths. For $t<0$ we approach it  along the isochemical potential line ($\hat\Phi=1$), whereas for $t>0$ we approach it 
	along the coexistence lines between   regions I and II. Due to our choice of
	various trajectories in the $\hat{Q}-\hat{\Phi}$ diagram, the
	discontinuity in the value of thermodynamic curvatures in Eq. \eqref{finalcritically} is unavoidable.
	Numerically checking this critical behaviour by plotting $R_{N}$ and $K_{N}$ versus $t$ (in a log-log plot), we summarize
the results for various cases in table  \ref{tab2}.
	
	\begin{table}[h]
		\begin{center}
			\begin{tabular}{c ||c||rrrrrr}
				\hline\hline
				Quantity & Coefficient & (11;4;3) & (10;5;0)& (11;7;-3) \\
				& & 3D CFT  & 4D CFT & 6D CFT \\\hline
				& $c_{R}$  & 2.00388 & 2.00567 & 2.0096 \\[-1ex]
				\raisebox{1.5ex}{$\ln |R_N|$ (Region (I))}
				& $d_{R}$ & 3.56833 & 4.54961 & 5.70467  \\\hline
				& $c_{R}$ & 1.99583 & 1.99386 & 1.99015  \\[-1ex]
				\raisebox{1.5ex}{$\ln |R_N|$ (Region (II))}&
				$d_{R}$ & 3.45195 & 4.37827 & 5.42222 \\\hline
				& $\c_{K}$ & 0.993519& 0.990341 & 1.01594  \\[-1ex]
				\raisebox{1.5ex}{$\ln |K_N|$ (Region (I))}&
				$d_{K}$ &2.00937& 2.44133 & 3.41461 \\\hline
				& $c_{K}$  &1.00561 & 1.00776 & 1.01143  \\[-1ex]
				\raisebox{1.5ex}{$\ln |K_N|$ (Region (II))}
				& $d_{K}$ &2.18417 & 2.2.69318 & 3.29687  \\ \hline\hline
			\end{tabular}
			\caption{The coefficients of the numerical fitting of $\ln|B_N|=-c_B\ln(t)-d_B \hspace{2mm} (B=R,K)$, for coexistence saturated Region (I) and coexistence saturated Region (II).}\label{tab2}
		\end{center}
	\end{table}
	
	Taking numerical error into account, the slopes of the  lines of the curves  $\ln|B_N|=-c_B\ln(t)-d_B \hspace{2mm} (B=R,K)$
in table \ref{tab2} imply that the critical exponents for $R_{N}$
	and $K_{N}$ are roughly $c_{R} = 2$ and $c_{K} = 1$, respectively,  consistent with those of charged AdS black holes in the bulk.  	
	\begin{table}[h]
		\begin{center}
			\begin{tabular}{c ||rrrrrr}
				\hline\hline
				Quantity &  (11;4;3) & (10;5;0)& (11;7;-3) \\
				& 3D CFT  & 4D CFT & 6D CFT \\\hline
				$R_{N} t^2$ &0.125214 & 0.125336 & 0.125153  \\\hline
				$\frac{(\beta-\gamma_{\tilde T})^2}{2}$  &0.125025 &0.12508  & 0.12528  \\\hline
				$K_{N} t$   & 0.251437 & 0.253177 & 0.25903 \\ \hline\hline
				$ \frac{(\gamma_{\tilde T}-\beta)}{2}$& 0.250025  & 0.25008 & 0.25028 \\ \hline\hline
			\end{tabular}
			\caption{Amplitudes of the critical behavior of normalized thermodynamic curvatures near the critical point.}\label{tab3}
		\end{center}
	\end{table}
From the values of $d_{R}$ and $d_{K}$ we collect the amplitudes related to these
	critical exponents  in table \ref{tab3}. We conclude  that the behavior of both thermodynamic curvatures is independent of the number of CFT dimensions. Indeed, they confirm universal amplitudes presented in Eq. (\ref{finalcritically}). On the gravity side, we saw previously that there is the same criticality behavior, i.e. Eq. \eqref{Adscriticality} for higher dimensional charged AdS black holes in such a way that it is independent of the spacetime dimension $d$ \cite{HosseiniMansoori:2020jrx}. Therefore, we can conclude that such  universality  holds for both bulk and boundary theories.

	\section{Conclusions} \label{con}
	
	The purpose of the current study was to apply the NTG geometry to determining the critical behavior and phase structure of the thermal boundary conformal field theories near the critical point. According to the AdS/CFT correspondence, such a $(d-1)$ dimensional boundary CFT is dual to a $d$- dimensional AdS black hole embedded in
	$D$-dimensional superstring/M-theory inspired models. 
	
	
	 In the framework of extended black hole thermodynamics \cite{Kubiznak:2012wp, Kastor:2009wy,Dolan:2011xt}, the cosmological constant is associated with the pressure of the gravitational system, $P=-\Lambda/8 \pi G$,
	whereas its conjugate quantity is interpreted as the thermodynamic volume $V$. The effect of including 
	these two variables in thermodynamic phase space has led to the discovery of a broad range of new
	phenomena associated with black holes \cite{gunasekaran2012extended,altamirano2013reentrant,hennigar2017superfluid,kubizvnak2017black}. Nonetheless, it is natural to ask what the interpretation is of the bulk pressure/volume on the dual boundary conformal field theory (CFT) once the cosmological constant is treated as a thermodynamic variable. To this end, there are two interpretations. As argued in \cite{Johnson:2014yja,Kastor:2014dra,Dolan:2014cja}, varying pressure, or
	$\Lambda$, is equivalent to varying the number of colors, $N$, in the boundary field theory so that the thermodynamic conjugate of pressure, i.e. the thermodynamic volume, can then be interpreted in the boundary field theory as an associated chemical potential, $\mu$  for color. Alternatively, the number of colors $N$  can be kept fixed, so that we are always referring to the same field theory, in which case varying
	$\Lambda$ in the bulk has the 
	more natural consequence of varying the volume of the space on which the field theory resides \cite{Karch:2015rpa}. 
	
	From this latter interpretation, we obtained critical exponents for heat capacities and compressibilities  in the $\hat{Q}-\hat{\Phi}$ plane with $\hat \Phi$ as the order parameter. The exponents are mean field, and the
	phase transition is
	similar to that of a Van der Waals gas, albeit one undergoing a reverse transition. Note that mean field exponents also
	characterize the phase transition in the $P-\mathcal{V}$ plane when $P$ would be the order parameter. 
	
	In addition, we studied the critical behavior of the normalized intrinsic and extrinsic thermodynamic curvatures near the critical point. More precisely, the criticality of such a curvature has been analytically and numerically checked form two trajectories, i.e., along the isochemical potential line and the coexistence curve in the $\hat{Q}-\hat{\Phi}$ diagram on approaching to the critical point. In analogy with charged AdS black hole cases in the bulk \cite{HosseiniMansoori:2020jrx}, our finding reveals that the critical exponent of the intrinsic and extrinsic curvature is 2 and 1, respectively. Furthermore, the analytical result shows a relation between the amplitude of thermodynamic curvatures and mean field critical exponents as shown in Eqs. \eqref{criticalRNS} and \eqref{criticalRNB}. Interestingly, critical amplitudes do not depend on the number of thermal CFT dimensions. Furthermore, we calculated numerically these amplitudes in Eq. \eqref{finalcritically}, which are in consistent with analytical results.
	Manifestly, there is always a universal feature of thermodynamic curvatures for both the bulk and boundary theories in the context of gauge/gravity duality.

	\vspace{0.5cm}
	\section*{Acknowledgements}
	We are grateful to Yu-Xiao Liu and Brian P. Dolan for
	reading a preliminary version of the draft.  This work was supported in part by the Natural Sciences and Engineering Research Council of Canada and the National Natural Science Foundation of China (Grant No. 12075103).
	\vspace{1cm}


\bibliography{references} 

\end{document}